\documentclass[conference]{IEEEtran}
\usepackage[T1]{fontenc}
\usepackage[latin1]{inputenc}
\usepackage{color}
\usepackage{psfrag}
\usepackage[dvips]{graphicx}
\usepackage{tikz}
\usepackage{pgfplots}
\pgfplotsset{compat=1.12}
\usepackage{enumerate}
\usepackage{rotating}
\usepackage{srcltx}
\usepackage{booktabs}
\usepackage{multicol}
\usepackage{enumitem}
\usepackage{graphicx}
\usetikzlibrary{arrows,decorations.pathmorphing,backgrounds,positioning,fit,matrix}

\usepackage[noadjust]{cite}
\usepackage[cmex10]{amsmath}
\usepackage{amsfonts}
\usepackage{amssymb}
\usepackage{ntheorem}
\usepackage{dsfont}    
\usepackage{mathtools} 
\usepackage{stackengine}

\definecolor{PU_orange}{HTML}{EE7F2D}
\definecolor{PU_darkorange}{HTML}{994400}
\definecolor{PU_lightorange}{HTML}{FFAA66}
\definecolor{PU_black}{HTML}{000000}
\definecolor{PU_darkgray}{HTML}{7F7F83}
\definecolor{PU_lightgray}{HTML}{BDBEC1}

\usepackage{algorithmic}
\usepackage{array}
\usepackage[caption=false,font=footnotesize]{subfig}

\theoremseparator{.}

\newtheorem{prop}{Proposition}
\newtheorem{lem}{Lemma}
\newtheorem{theorem}{Theorem}
\theoremseparator{.}

\newtheorem{remark}{Remark}
\newtheorem*{example*}{Example}

\newcommand{\E}{\mathbb{E}}

\long\def\symbolfootnote[#1]#2{\begingroup%
	\def\thefootnote{\fnsymbol{footnote}}\footnote[#1]{#2}\endgroup} 
\title{Measuring Dependencies of Order Statistics: 
\\An Information Theoretic Perspective
}%
\author{
\IEEEauthorblockN{Alex Dytso$^{\star}$, Martina Cardone$^*$, Cynthia Rush$^\dagger$}
$^{\star}$ New Jersey Institute of Technology, Newark, NJ 07102, USA Email: alex.dytso@njit.edu\\
$^*$ University of Minnesota, Minneapolis, MN 55404, USA, Email: cardo089@umn.edu\\
$^\dagger$ Columbia University, New York, NY 10025, USA, Email: cynthia.rush@columbia.edu
\vspace{-0.9em}
\thanks{The work of M. Cardone was supported in part by the U.S. National Science Foundation under Grant CCF-1849757. 
}
}

\begin{document}
\IEEEoverridecommandlockouts
\maketitle

\begin{abstract}
Consider a random sample $X_1 , X_2 , ..., X_n$ drawn independently and identically distributed from some known sampling distribution $P_X$. Let $X_{(1)} \le X_{(2)} \le ... \le X_{(n)}$ represent the order statistics of the sample. 
The first part of the paper focuses on distributions with an invertible cumulative distribution function.  Under this assumption, a distribution-free property is established, which shows that the $f$-divergence between the joint distribution of order statistics and the product distribution of order statistics does not depend on the original sampling distribution $P_X$. Moreover, it is shown that the mutual information between two subsets of order statistics also satisfies a distribution-free property; that is, it does not depend on $P_X$. Furthermore, the  decoupling rates between $X_{(r)}$ and $X_{(m)}$ (i.e., rates at which the mutual information approaches zero) are characterized for various choices of $(r,m)$.    
The second part of the paper considers a family of discrete distributions, which does not satisfy the assumptions in the first part of the paper.  In comparison to the results of the first part, it is shown that in the discrete setting, the mutual information between order statistics does depend on the sampling distribution $P_X$.  Nonetheless, it is shown that the results of the first part can still be used as upper bounds on the 
decoupling rates.
\end{abstract}

\section{Introduction}
Consider a random sample $X_1 , X_2 , ..., X_n$ drawn independently and identically distributed (i.i.d.) from some known  sampling distribution $P_X$.  Let the random variables $X_{(1)} \le X_{(2)}  \le ... \le X_{(n)}$  represent the order statistics of the sample.
In this work, we are interested in studying  the dependence between 
$X_{(\mathcal{I}_1)}$ and $X_{(\mathcal{I}_2)}$ where $\mathcal{I}_1$ and $\mathcal{I}_2$ are two arbitrary subsets of $\{1,...,n\}$ and $X_{(\mathcal{I}_k)}= \{ X_{(i)} \}_{i \in \mathcal{I}_k},$ for $k \in \{1,2 \}$. In particular, we choose to use the $f$-divergence and mutual information as measures of such dependence.

Our contributions and paper outline are as follows.
 In Section~\ref{sec:Cont_Dist}, we consider
the
$f$-divergence and the mutual information of order statistics when the sample is drawn from a large family of distributions, namely, the set of all distributions having an invertible cumulative distribution function (cdf).  Under this assumption, we show that the$f$-divergence between the joint distribution and the product distribution does not depend on the sampling distribution. Moreover, under this distributional assumption, for every finite $n$, 
we show
that the mutual information between $X_{(\mathcal{I}_1)}$ and $X_{(\mathcal{I}_2)}$
does not depend on the sampling distribution, and we compute the exact value of the mutual information  for the case $\mathcal{I}_1 = \{r\}$ and $\mathcal{I}_2 = \{m\}$ for integers $1 \leq r < m \leq n$.  
Furthermore,  we characterize the rates of decoupling  between $X_{(r)}$  and $X_{(m)}$ (i.e., rates at which the mutual information approaches zero)  
for various choices of  $(r,m)$.      For example, 
we show
that the minimum and maximum (i.e., $(r,m)=(1,n)$) decouple at a rate of $\frac{1}{n^2}$ while the median and maximum decouple at a rate of $\frac{1}{n}$. 
In Section~\ref{sec:Discrete_Case}, we consider a family of discrete distributions, which does not fall into the family 
of
Section~\ref{sec:Cont_Dist}. In comparison to the results in Section~\ref{sec:Cont_Dist}, we show that in the discrete setting, the mutual information between $X_{(r)}$  and $X_{(m)}$ does depend on the sampling distribution.   Nonetheless, we prove that the results in Section~\ref{sec:Cont_Dist} can still be used as upper bounds on the decoupling rates in the discrete setting. Finally, to provide some comparisons, we compute the mutual information between $X_{(r)}$  and $X_{(m)}$ for the case when the sampling distribution comes from the Bernoulli distribution. 

\smallskip
\noindent
{\bf Related Work.}
Order statistics have a wide range of applications in statistical signal processing; the interested reader is referred to~\cite{HBS17} for a comprehensive survey. 
Information measures of the distribution of order statistics have also received some attention.  For example, the authors of \cite{baratpour2007some} showed conditions under which the differential entropy of the order statistics characterizes the sampling distribution. 
Other information measures that have been considered on the distribution of order statistics include the R\'enyi entropy~\cite{baratpour2008characterizations,abbasnejad2010renyi},   the cumulative entropies \cite{balakrishnan2020cumulative}, and the Fisher information~\cite{zheng2009fisher}. 

Distribution-free properties for information measures on order statistics have also been observed in the past.
For instance,
the authors of~\cite{abbasnejad2010renyi}, for continuous distributions, have shown that the R\'enyi divergence between order statistics and their sampling distribution does not depend on the underlying sampling distribution.  
The authors of~\cite{wong1990entropy}, for continuous distributions, have shown that the average entropy of the individual order statistics and the entropy of the sampling distribution do not depend on the underlying sampling distribution.  The authors of \cite{ebrahimi2004information}, for continuous distributions, have shown that the mutual information between consecutive order statistics is independent of the sampling distribution provided.

We generalize the above results in several directions. First, we show a distribution-free property for the $f$-divergence. Second, the proof technique that we use allows us to extend the distribution-free property beyond continuous distributions and to arbitrary subsets of random variables. Third, we find the exact large $n$ behavior of the mutual information between $X_{(r)}$ and $X_{(m)}$ for various regimes. 

\smallskip
\noindent
{\bf Notation.}
We use $[n]$ to denote the collection $\{1, 2,..., n\}$. Logarithms are assumed to be in base $e$.
The notation $\stackrel{D}{=}$ denotes equality in distribution.
The harmonic number, denoted as $H_r$, is defined as follows. For $r\in \mathbb{N}$,
\begin{equation}
\label{eq:Hr_def}
H_r= \sum_{k=1}^r \frac{1}{k}. 
\end{equation} 
We also define, for $r\in \mathbb{N}$,
\begin{equation}
\label{eq:Tr_def}
T_r = \log(r!)- rH_r. 
\end{equation}
 The  Euler-Mascheroni constant is denoted by  $\gamma\approx 0.5772$.    
 Let $f: (0, \infty) \to \mathbb{R}$ be a convex function such that $f(0)=1$. Then, for two probability distributions $P$ and $Q$  over a space $\Omega$ such that $P \ll Q$ (i.e., $P$ is absolutely continuous with respect to $Q$), the $f$-divergence is defined as
 \begin{equation}
 D_f(P\|Q)=  \int_{\Omega} f \left( \frac{{\rm d}P}{{\rm d} {\rm}Q} \right) {\rm d} Q. 
 \end{equation}

\section{The Case of Continuous Distributions} 
\label{sec:Cont_Dist}
In this section, we consider a setting in which the cdf of a sampling distribution is an invertible function (i.e., bijective function).  
Several classes of probability distributions satisfy this property. 
For example, all absolutely continuous distributions with a non-zero probability density functions (pdf) satisfy this property since, in this case, the cdfs are strictly increasing and, therefore, have an inverse. 
A non-example, however, is the set of discrete distributions having step functions for their cdfs, which do not have a proper inverse. 

Out of the two aforementioned classes of distributions,  the class of distributions with a non-zero pdf is one that is typically studied the most in conjunction with the order statistics \cite{nagaraja1982non}. Because the probability of ties in the sample for this case equals zero,  the analysis  considerably simplifies. For discrete distributions, one must account for the possibility of samples taking the exact same value, and the analysis often becomes combinatorially cumbersome. Nonetheless, we consider the case of discrete distributions  in the next section.

\subsection{Distribution-Free Property for the $f$-divergence}  
We begin our study on the dependence structure of order statistics by showing  that a large class of divergences, namely the $f$-divergence, have the following distribution-free property: if the cdf of the sampling distribution is invertible, then  the $f$-divergence between the joint distribution of order statistics  and the product distribution of order statistics  does not depend on the sampling distribution.  
\begin{theorem}\label{thm:f-diverngece} Fix a subset $\mathcal{I} \subseteq [n]$ and assume that $X_1, ..., X_n$ i.i.d.\ $\sim P_X$, with $P_X$ having an invertible cdf. Then,
\begin{equation}
D_f \Big(  P_{ \{ X_{(i)} \}_{i \in \mathcal{I}}  } \, \Big \|    \, \prod_{i \in \mathcal{I} }  P_{ X_{(i)} }  \Big) \hspace{-0.025cm}= \hspace{-0.025cm} D_f \Big( P_{ \{ U_{(i)} \}_{i \in \mathcal{I}}  } \, \Big  \| \,  \prod_{i \in \mathcal{I} }  P_{ U_{(i)} } \Big), \label{eq:f-divergence}
\end{equation} 
where $P_{ \{ X_{(i)} \}_{i \in \mathcal{I}}  }$ and $\prod_{i \in \mathcal{I} }  P_{ X_{(i)} }$ are the joint distribution and  the product distribution of the sequence  $\{ X_{(i)} \}_{i \in \mathcal{I}}  $, respectively;  
$(U_{(1)}, ..., U_{(n)})$ are the order statistics associated with the 
sample $(U_{1}, ..., U_{n})$ i.i.d.\ $\sim \mathcal{U}(0,1)$, where $\mathcal{U}(0,1)$ denotes the uniform distribution over $(0,1)$; and $P_{ \{ U_{(i)} \}_{i \in \mathcal{I}}}  $ and $\prod_{i \in \mathcal{I} }  P_{ U_{(i)} } $  are the joint distribution and  the product distribution of the sequence  $\{ U_{(i)} \}_{i \in \mathcal{I}}  $, respectively. 
\end{theorem} 
\begin{IEEEproof}
Let $F_X^{-1}$ be the inverse cdf of the sampling distribution $P_X$. Recall that for $(U_{1}, ..., U_{n})$ i.i.d.\ $\sim \mathcal{U}(0,1)$,
we have $(X_1, ..., X_n)  \stackrel{D}{=} ( F_X^{-1}(U_{1}),  ..., F_X^{-1}(U_{n}))$. Then since $F_X^{-1}(\cdot)$ is order preserving (see~\cite[eq.\ (2.4.2)]{arnold1992first}), we have
\begin{align}
X_{(\mathcal{I})} \, = \, \{ X_{(i)} \}_{i \in \mathcal{I}} \, \stackrel{D}{=} \, \{  F_X^{-1}(U_{(i)}) \}_{i \in \mathcal{I}} .
\end{align} 
Then since $F_X$ is a one-to-one mapping  and the $f$-divergence   is invariant under invertible transformations \cite[Thm.~14]{liese2006divergences}, 
\begin{align}
&D_f \Big(  P_{ \{ X_{(i)} \}_{i \in \mathcal{I}}  } \, \Big \|  \,  \prod_{i \in \mathcal{I} }  P_{ X_{(i)} } \Big)\notag\\
&=D_f \Big(  P_{ \{ F_X^{-1}(U_{(i)}) \}_{i \in \mathcal{I}}  } \, \Big \| \,  \prod_{i \in \mathcal{I} }  P_{ F_X^{-1}(U_{(i)}) }  \Big) \notag\\
&= D_f \Big(  P_{ \{ U_{(i)} \}_{i \in \mathcal{I}}  } \, \Big \| \,  \prod_{i \in \mathcal{I} }  P_{ U_{(i)} } \Big). 
\end{align} 
This concludes the proof of Theorem~\ref{thm:f-diverngece}.
\end{IEEEproof} 
\begin{remark} Computing the $f$-divergence in \eqref{eq:f-divergence} requires the knowledge of the joint distribution of $\{ U_{(i)} \}_{i \in \mathcal{I}}$ for any subset $\mathcal{I}$.  The joint  pdf of this sequence can be readily computed and is given by  the following expression~\cite{arnold1992first}: let $\mathcal{I}=\{ (i_1, i_2,..., i_k): 1 \le i_1, i_2,..., i_k \le n  \} $ where $|\mathcal{I}|=k$, then, $P_{ \{ U_{(i)} \}_{i \in \mathcal{I}}} $
is non-zero only if
$
-\infty < x_{(i_1)} < x_{(i_2)} < ... < x_{(i_k)} < \infty,
$
and, when this is true, its expression is
\begin{align}
P_{ \{ U_{(i)} \}_{i \in \mathcal{I}}} 
= c_{\mathcal{I}}
 \prod_{t=1}^{k+1} \left [x_{(i_t)} - x_{(i_{t-1})} \right ]^{i_t-i_{t-1}-1}, \label{eq:Joint_Distribution}
\end{align}
where $x_{(i_0)} = x_{(i_{k+1})} =0$, and, with $i_0 = 0$ and $i_{k+1} = n+1$,
\begin{align*}
c_{\mathcal{I}} = \frac{n!}{\prod_{t=1}^{k+1} (i_t - i_{t-1}-1)!}.
\end{align*}
\end{remark}
The next result, the proof of which is in Appendix~\ref{sec:Computation_Of_KL}, evaluates the 
Kullback-Leibler (KL) divergence, which is a special case of the $f$-divergence with $f(x)=x \log(x)$. 
\begin{prop}\label{prop:KL_div} Under the assumptions of Theorem~\ref{thm:f-diverngece}, where $\mathcal{I} \subseteq [n]$ with $|\mathcal{I}|=k$, we have that 
\begin{align}
&D_{\text{KL}} \Big(  P_{ \{ U_{(i)} \}_{i \in \mathcal{I}}  }  \, \Big \|  \, \prod_{i \in \mathcal{I} }  P_{ U_{(i)} }  \Big) \notag\\
& =  \sum_{t=2}^k (T_{i_t-1} -T_{i_t-i_{t-1}-1})+\sum_{t=1}^{k-1} T_{n-i_t}-(k-1)T_n. \label{eq:KL_expression}
\end{align}
In particular, \\
\textbf{ (Whole Sequence).} For $\mathcal{I}=[n]$, we have that 
\begin{align*}
&D_{\text{KL}} \Big(  P_{ \left \{U_{(1)}, ...,  U_{(n)} \right \} } \, \Big \| \,  \prod_{i=1}^n  P_{ U_{(i)} } \Big)  = 2 \sum_{t=2}^n T_{t-1} -(n-1)T_n.
\end{align*} \\
\textbf{ (Min and Max). }  For $\mathcal{I}=\{1, n\}$, we have that 
\begin{align*}
&D_{\text{KL}}\Big(  P_{ \left \{U_{(1)}, \,  U_{(n)} \right \}} \, \Big \| \,   P_{ U_{(1)} }  \,P_{ U_{(n)} }  \Big)  
=  \log \left( \frac{n-1}{n} \right ) + \frac{1}{n-1}.
\end{align*}
\end{prop}

\begin{remark}
For $\mathcal{I}=\{1, n\}$,  Proposition~\ref{prop:KL_div} says that, when $n \to \infty$, we have that
\begin{align*}
&\lim_{n \to \infty} n^2 D_{\text{KL}}\Big(  P_{ \left \{U_{(1)}, \,  U_{(n)} \right \}} \, \Big \| \,   P_{ U_{(1)} }  \,P_{ U_{(n)} }  \Big)  \\
& \qquad =\lim_{n \to \infty} n^2 \left [  \log \left( \frac{n-1}{n} \right ) + \frac{1}{n-1} \right ] =  \frac{1}{2},
\end{align*}
where the last equality follows by using the Maclaurin series for the natural logarithm. Thus, when the KL divergence is considered, the joint and product distributions of the minimum and maximum converge at a rate equal to $1/n^2$.
\end{remark}

\subsection{Distribution-Free Property for the Mutual Information}
Here we consider the mutual information measure. 
In particular,
as a special case of the approach used for the proof of Theorem~\ref{thm:f-diverngece} we have the following result.
\begin{theorem}\label{thm:Exact_Char_MI_Order_Stat}
Assume that $X_1, ..., X_n$ i.i.d.\ $\sim P_X$, with $P_X$ having an invertible cdf and fix two sets  $\mathcal{I}_1, \mathcal{I}_2 \subseteq [n]$.  Then, 
\begin{equation}
I(X_{(\mathcal{I}_1)}; X_{(\mathcal{I}_2)}) = I(U_{(\mathcal{I}_1)}; U_{(\mathcal{I}_2)}), \label{eq:MI_continious_case}
\end{equation} 
where $X_{(\mathcal{I}_k)}= \{ X_{(i)} \}_{i \in \mathcal{I}_k}$ and $U_{(\mathcal{I}_k)}= \{ U_{(i)} \}_{i \in \mathcal{I}_k}$ both for $k \in \{1,2 \}$.  Consequently,   $I(X_{(\mathcal{I}_1)}; X_{(\mathcal{I}_2)})$ is \textbf{not a function} of $P_X$, the sampling distribution of $X$. 
Moreover, for $r<m$ 
\begin{align}
I(X_{(r)}; X_{(m)}) = T_{m-1} + T_{n-r} - T_{m-r-1} - T_n.
\label{eq:Exact_MI_Continious}
\end{align}  \end{theorem}

\begin{IEEEproof}
The proof of~\eqref{eq:MI_continious_case} follows along the same lines as the proof of Theorem~\ref{thm:f-diverngece} and relies on the invariance of the mutual information to one-to-one transformations. 

To compute \eqref{eq:Exact_MI_Continious}, recall that the  mutual information can be written as a KL divergence, and then using  \eqref{eq:KL_expression},
 \begin{align*}
 I(U_{(r)}; U_{(m)})&= D_{\text{KL}}(P_{U_{(r)}, U_{(m)}} \| P_{U_{(r)}} P_{ U_{(m)}} )  \\
 &= T_{m-1} + T_{n-r} - T_{m-r-1} - T_n.
 \end{align*}
This concludes the proof of Theorem~\ref{thm:Exact_Char_MI_Order_Stat}.
\end{IEEEproof}

\begin{remark}  In Theorem~\ref{thm:Exact_Char_MI_Order_Stat},  if $\mathcal{I}_1 \cap \mathcal{I}_2 \neq \varnothing $, then  $I(X_{(\mathcal{I}_1)}; X_{(\mathcal{I}_2)})=\infty$. 
Moreover, the assumption  that $r<m$  is without loss of generality since the mutual information is symmetric.   \end{remark}

\begin{remark}

For other measures of dependence of random variables, the distribution independence property of Theorem~\ref{thm:Exact_Char_MI_Order_Stat} does not necessarily hold.   For example, as demonstrated in Appendix~\ref{app:eq:cov_1_2_exponential}, the covariance of a pair of order statistics from a sample drawn according to an exponential distribution with rate $\lambda$ is
\begin{equation}
{\rm Cov}(X_{(1)},X_{(2)}) = {1}/{(\lambda^2 n^2)}. \label{eq:cov_1_2_exponential}
\end{equation}
\end{remark}

The Theorem~\ref{thm:Exact_Char_MI_Order_Stat} eq.\ \eqref{eq:Exact_MI_Continious} result is stated in terms of factorials and harmonic numbers, as captured by the $T$'s defined in~\eqref{eq:Tr_def}. 
In the following lemma (see Appendix~\ref{app:lem:approx_lemma} for the proof), we provide  an alternative formulation for these $T$'s that will be helpful for the large sample size analysis. 
\begin{lem}\label{lem:approx_lemma} For $k >0$,
\begin{align}
&T_k = k\hspace{-0.05cm} \log \hspace{-0.05cm} \left( \hspace{-0.05cm}\frac{2k}{2k+1} \hspace{-0.05cm}\right) + \frac{1}{2} \log(2 \pi k) -(1+\gamma) k  - e(k),\label{eq:Tk_approx} \\
 &T_{k+1} - T_{k} = \log \hspace{-0.05cm} \left( \hspace{-0.05cm}\frac{2k+2}{2k+3}\hspace{-0.05cm}\right) -(1+\gamma)  +\frac{1}{k+1} -c(k), \label{eq:diff_of_T(n)-T(n-1)}
\end{align} 
where 
\begin{align}
\frac{k}{24 (k+1)^2}-\frac{1}{12k} &\le   e(k) \le    \frac{1}{24 k}-\frac{1}{12k+1}, \label{eq:e_approx} \\
\frac{1}{24 (k+2)^2}  &\le  c(k) \le   \frac{1}{24 (k+1)^2}.  \label{eq:ck}
\end{align} 
\end{lem}

\subsection{Large Sample Size Asymptotics of Mutual Information} 
\begin{figure}
\center
%
%
\definecolor{mycolor1}{rgb}{0.85000,0.32500,0.09800}%
\begin{tikzpicture}

\begin{axis}[%
width=7cm,
height=2.8cm,
at={(1.011in,0.642in)},
scale only axis,
xmin=0,
xmax=100,
xlabel style={font=\color{white!15!black}},
xlabel={$n$},
ylabel={ \small $ n I(X_{(\lfloor  \frac{n}{2} \rfloor)}; X_{(n)})$},
ymin=0.2,
ymax=0.7,
axis background/.style={fill=white},
xmajorgrids,
ymajorgrids,
legend style={legend cell align=left, align=left, draw=white!15!black}
]
\addplot [color=black, thick]
  table[row sep=crcr]{%
2	0.613705638880109\\
3	0.283604675675505\\
4	0.560744611093554\\
5	0.362538547836722\\
6	0.541116916640338\\
7	0.399356151118711\\
8	0.531013031710899\\
9	0.420634301595234\\
10	0.524877400749713\\
11	0.434490287710609\\
12	0.520761971808781\\
13	0.444229107457875\\
14	0.517812044033434\\
15	0.451447864242009\\
16	0.515594716990506\\
17	0.457012422081625\\
18	0.513867586025825\\
19	0.461432823057855\\
20	0.512484452309678\\
21	0.46502900325806\\
22	0.511351945159205\\
23	0.468011801572011\\
24	0.510407655589461\\
25	0.470525803404165\\
26	0.50960829355455\\
27	0.472673466689606\\
28	0.508922891634626\\
29	0.474529422074177\\
30	0.508328713943982\\
31	0.476149318827041\\
32	0.507808692722392\\
33	0.477575498308909\\
34	0.507349766320601\\
35	0.478840747399936\\
36	0.506941769382053\\
37	0.47997084945915\\
38	0.506576672484925\\
39	0.480986359466009\\
40	0.506248049309477\\
41	0.481903865282604\\
42	0.505950695677711\\
43	0.482736900666524\\
44	0.505680352284401\\
45	0.483496617178147\\
46	0.505433499609609\\
47	0.484192286247833\\
48	0.50520720402892\\
49	0.484831679340473\\
50	0.5049990007997\\
51	0.485421359526413\\
52	0.504806803966744\\
53	0.485966907536778\\
54	0.504628836343102\\
55	0.486473098917344\\
56	0.50446357438932\\
57	0.486944044218774\\
58	0.504309704547666\\
59	0.487383300816148\\
60	0.504166088281579\\
61	0.487793962939506\\
62	0.504031733844982\\
63	0.488178734682478\\
64	0.503905773396582\\
65	0.488539989603396\\
66	0.503787444204761\\
67	0.488879819742408\\
68	0.503676073215615\\
69	0.489200076142652\\
70	0.503571064283221\\
71	0.489502402556241\\
72	0.503471887451724\\
73	0.489788263741104\\
74	0.503378070022336\\
75	0.490058969106144\\
76	0.503289189030852\\
77	0.490315692823415\\
78	0.503204864884168\\
79	0.490559490835096\\
80	0.503124755932731\\
81	0.490791315322497\\
82	0.503048553849794\\
83	0.491012027227669\\
84	0.502975979638222\\
85	0.491222406884191\\
86	0.502906780270735\\
87	0.491423163204843\\
88	0.502840725725036\\
89	0.491614941724094\\
90	0.502777606361633\\
91	0.491798331523171\\
92	0.502717230809822\\
93	0.491973871176782\\
94	0.502659423997272\\
95	0.492142054208529\\
96	0.502604025416986\\
97	0.492303333559647\\
98	0.502550887628843\\
99	0.492458125765637\\
100	0.502499875031504\\
};
\addlegendentry{Exact in \eqref{eq:MI_continious_case}}

\addplot [color=red, dashed, thick]
  table[row sep=crcr]{%
1	0.5\\
2	0.5\\
3	0.5\\
4	0.5\\
5	0.5\\
6	0.5\\
7	0.5\\
8	0.5\\
9	0.5\\
10	0.5\\
11	0.5\\
12	0.5\\
13	0.5\\
14	0.5\\
15	0.5\\
16	0.5\\
17	0.5\\
18	0.5\\
19	0.5\\
20	0.5\\
21	0.5\\
22	0.5\\
23	0.5\\
24	0.5\\
25	0.5\\
26	0.5\\
27	0.5\\
28	0.5\\
29	0.5\\
30	0.5\\
31	0.5\\
32	0.5\\
33	0.5\\
34	0.5\\
35	0.5\\
36	0.5\\
37	0.5\\
38	0.5\\
39	0.5\\
40	0.5\\
41	0.5\\
42	0.5\\
43	0.5\\
44	0.5\\
45	0.5\\
46	0.5\\
47	0.5\\
48	0.5\\
49	0.5\\
50	0.5\\
51	0.5\\
52	0.5\\
53	0.5\\
54	0.5\\
55	0.5\\
56	0.5\\
57	0.5\\
58	0.5\\
59	0.5\\
60	0.5\\
61	0.5\\
62	0.5\\
63	0.5\\
64	0.5\\
65	0.5\\
66	0.5\\
67	0.5\\
68	0.5\\
69	0.5\\
70	0.5\\
71	0.5\\
72	0.5\\
73	0.5\\
74	0.5\\
75	0.5\\
76	0.5\\
77	0.5\\
78	0.5\\
79	0.5\\
80	0.5\\
81	0.5\\
82	0.5\\
83	0.5\\
84	0.5\\
85	0.5\\
86	0.5\\
87	0.5\\
88	0.5\\
89	0.5\\
90	0.5\\
91	0.5\\
92	0.5\\
93	0.5\\
94	0.5\\
95	0.5\\
96	0.5\\
97	0.5\\
98	0.5\\
99	0.5\\
100	0.5\\
};
\addlegendentry{Limit in \eqref{eq:alpha_max_limit_n}}

\end{axis}

\end{tikzpicture}%
\vspace{-3mm}
\caption{Convergence of $n I(X_{(\lfloor  \frac{n}{2} \rfloor)}; X_{(n)}) $  (Median vs. Max) to~\eqref{eq:Median_max_limit_n}. }
\label{fig:Median_MI_n}
\vspace{-0.5cm}
\end{figure}
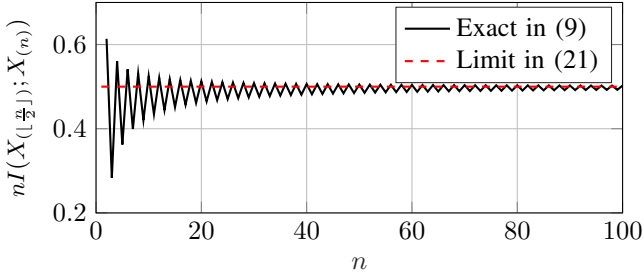

Using Theorem~\ref{thm:Exact_Char_MI_Order_Stat} and the approximations in Lemma~\ref{lem:approx_lemma}, we now study the rates  of decoupling of the order statistics as the sample size grows. 
In particular, we have the next theorem, which is proved in Appendix~\ref{app:thm:Limits_Continious_Case}.
\begin{theorem}\label{thm:Limits_Continious_Case}  Under the assumptions of Theorem~\ref{thm:Exact_Char_MI_Order_Stat}, we have the following: 
\begin{enumerate}[leftmargin=*]
\item  ($r^{th}$ vs.\ Max). Fix some $r \ge 1$ independent of $n$. Then,  
\begin{equation}
\lim_{n \to \infty}  n^2 I(X_{(r)}; X_{(n)})={r}/{2}.  \label{eq:r-th_order_vs_max}
\end{equation} 
\item  ($r^{th}$ vs.\ $m^{th}$).
Fix some $1 \leq r<m$ independent of $n$. Then, 
\begin{equation}
\lim_{n \to \infty}   I(X_{(r)}; X_{(m)}) = T_{m-1} - T_{m-r-1}+(1+\gamma) r.  \label{eq:Fixed_m_r_infinity_n}
\end{equation} 
\item    ($k$-Step).  Fix some $k \ge 1$. Then, 
\begin{equation}
\begin{split}
&\lim_{n \to \infty}   I(X_{(n-k)}; X_{(n)}) =  \log(k)  -H_{k-1} +\gamma \\
&\qquad \qquad =\log \left(\frac{k}{k+\frac{1}{2}}\right)+\frac{1}{k} -c(k).  \label{eq:k-step_limit}
\end{split}
\end{equation} 
Consequently, for $1$ step,
\begin{equation} \label{eq:1step}
\lim_{n \to \infty}   I(X_{(n-1)}; X_{(n)})= \gamma . 
\end{equation} 
\item  ($\lfloor \alpha n \rfloor$ vs.\ $\lceil \beta n \rceil $). Fix $0<\alpha <\beta < 1$, with $(\alpha,\beta)$ independent of $n$. Then, 
\begin{equation}
\lim_{n \to \infty}   I(X_{(\lfloor \alpha n \rfloor)}; X_{(\lceil \beta n \rceil )})=\frac{1}{2} \log \left( \frac{\beta (1-\alpha)}{\beta-\alpha} \right). \label{eq:Beta-alpha_limit_MI}
\end{equation} 
\item  ($\lfloor \alpha n \rfloor$ vs. Max).   Fix some  $0<\alpha<1$ with $\alpha$ independent of $n$. Then,  
\begin{equation}
\lim_{n \to \infty}  n I(X_{(\lfloor \alpha n \rfloor)}; X_{(n)}) = \frac{\alpha}{2(1-\alpha)}.  \label{eq:alpha_max_limit_n}
\end{equation} 
Consequently,  
we have the following limits:
\begin{align}
\text{ $Q_3$ vs. Max: }& \lim_{n \to \infty}   n I(X_{(\lfloor  \frac{ 3n}{4} \rfloor)}; X_{(n)}) =\frac{3}{2},\\
 \text{ Median vs. Max: }&\lim_{n \to \infty}   n I(X_{(\lfloor  \frac{n}{2} \rfloor)}; X_{(n)}) =\frac{1}{2}, \label{eq:Median_max_limit_n} \\
  \text{ $Q_1$ vs. Max: }& \lim_{n \to \infty}   n I(X_{(\lfloor  \frac{n}{4} \rfloor)}; X_{(n)}) =\frac{1}{6}. 
\end{align} 
\end{enumerate}
\end{theorem} 
For comparison, Fig.\ \ref{fig:Median_MI_n} demonstrates how the median decouples from the maximum for finite values of $n$.  
\begin{remark}
We compare the rate of decoupling of the mutual information $I(U_{(r)}; U_{(m)})$ for integers $r < m$ to that of the covariance between $U_{(r)}$ and $U_{(m)}$ given by~\cite{arnold1992first},
\begin{align}
{\rm Cov} (U_{(r)},U_{(m)} ) =\frac{r (n-m+1)}{(n+1)^2 (n+2)}.   \label{eq:Order_covariance_Unfiorm_r_m} 
\end{align} 
Note that, although this comparison is somewhat unfair as the covariance only captures correlation, it can still be used as  a proxy for measuring independence. 
From Table~\ref{table:Comparison}, we observe that the rates of decoupling of the mutual information and covariance are always different. 
Moreover, we also note a surprising behavior for Cases~2-4: although the mutual information does not decouple, the covariance goes to zero either at a rate $1/n^2$ (Cases~2-3) or at a rate $1/n$ (Case~4).
Finally, by comparing Case~4 and Case~5, we observe that there is a phase transition at $\beta=1$, i.e., in Case~4 ($0<\beta<1$) the mutual information does not decouple, whereas in Case~5 ($\beta=1$) it decouples at a rate $1/n$.
\end{remark}
\begin{table}[ht]
\centering
\begin{tabular}[t]{lcc}
\hline
& ${\rm Cov} (U_{(r)},U_{(m)} )$ & $I(U_{(r)}; U_{(m)})$\\
\hline
Case~1: $m=n$ & $1/n^3$ & $1/n^2$\\
Case~2 & $1/n^2$ & No decoupling \\
Case~3: $r=n-k$ and $m=n$ & $1/n^2$ & No decoupling \\
Case~4: $r=\lfloor \alpha n \rfloor$ and $m=\lceil \beta n \rceil $ & $1/n$ & No decoupling \\
Case~5: $r=\lfloor \alpha n \rfloor$ and $m=n$ & $1/n^2$ & $1/n$ \\
\hline
\end{tabular}
\vspace{2mm}
\caption{Rates of decoupling for ${\rm Cov} (U_{(r)},U_{(m)} )$ and $I(U_{(r)}; U_{(m)})$.}
\label{table:Comparison}
\vspace{-5mm}
\end{table}

\section{The Case of Discrete Distributions} 
\label{sec:Discrete_Case}

In this section, we consider the case when the sampling distribution is discrete. Historically, order statistics with a discrete distribution have received far less attention than those with a continuous distribution.  However, recently,  since  discrete distributions naturally occur in several practical situations (e.g., image processing), discrete order statistics have  started to receive more attention in the literature~\cite{evans2006distribution,dembinska2014discrete}.

The mutual information of discrete order statistics often behaves differently from that of continuous order statistics. For example, while order statistics $X_{(1)} \le X_{(2)}  \le ... \le X_{(n)}$ from a continuous distribution form a Markov chain~\cite{arnold1992first},
for the case of discrete distributions, the order statistics form a Markov chain if and only if the sampling distribution has at most two points in its support~\cite{nagaraja1982non}.

\begin{figure}
\center
\input{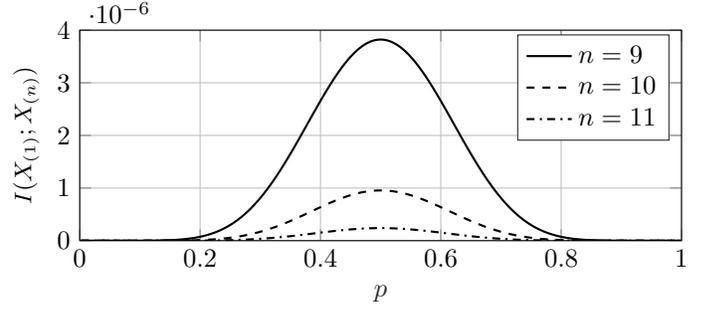}
\vspace{-8mm}
\caption{$I(X_{(1)}; X_{(n)} ) $ in \eqref{eq:min_vs_max_Bernoulli} versus $p \in (0,1)$.  }
\label{fig:MI_vs_p_Bernouilli}
\vspace{-0.5cm}
\end{figure}

The next theorem (see Appendix~\ref{app:thm:Bernoulli_Case} for the proof) shows that, unlike in the case of continuous order statistics, when the sampling distribution is discrete, the mutual information between $X_{(r)}$ and  $X_{(m)}$ can indeed depend on the sampling distribution.  
\begin{theorem}\label{thm:Bernoulli_Case} Suppose that  the sampling distribution is Bernoulli with parameter $p \in (0,1)$.  Then,   for $r \le m$,
\begin{align}
&I(X_{(r)}; X_{(m)} ) = - P( B \ge m  ) \log \left(  P( B \ge r  )   \right) \notag \\
&+\! (P( B \ge r) \!-\! P( B \ge m))\!   \log \left(\! \frac{P( B \ge r) \!-\! P( B \ge m)}{ P( B \ge r  )  (1\!-\!P( B \ge m ))} \!\right ) \notag\\
&-(1- P( B \ge r  ))  \log \left(1-P( B \ge m ) \right) ,\label{eq:MI_Bernoulli}
\end{align} 
where $B$ is  ${\rm Binomial}(n, 1-p) $.    Consequently, 
\begin{align}
&I(X_{(1)}; X_{(n)} ) 
\notag\\
&= -(1-p)^n \log \left( 1-p^n \right) - p^n  \log \left(  1-(1-p)^n \right)  \notag\\
&+ (1\!-\!p^n\!-\!(1-p)^n) \log \left( \frac{ 1-p^n- (1-p)^n}{ (1-p^n)  (1-(1-p)^n)} \right).
  \label{eq:min_vs_max_Bernoulli}
\end{align} 
\end{theorem}
Theorem~\ref{thm:Bernoulli_Case} provides an example of discrete sampling distribution (i.e., Bernoulli with parameter $p \in (0,1)$) for which the mutual information \emph{does} depend on the sampling distribution (i.e., parameter $p$).
For illustration, Fig.~\ref{fig:MI_vs_p_Bernouilli} shows the dependence of $I(X_{(1)}; X_{(n)} )$ in~\eqref{eq:min_vs_max_Bernoulli} over~$p \in (0,1)$. 

The fact that the mutual information does depend on the sampling distribution prevents us  from having universal results similar to those derived in Section~\ref{sec:Cont_Dist}. 
However, the results in Section~\ref{sec:Cont_Dist} can still be used as upper bounds as we show next. 

\begin{theorem}\label{thm:UpperBOund} Assume that $X_1, ..., X_n$ i.i.d.\ $\sim P_X$, with $P_X$ having an arbitrary sampling distribution.  Then, 
\begin{itemize}[leftmargin=*]
\item  (f-divergence). For   $\mathcal{I} \subset [n]$,
\begin{equation}
D_f \Big(  P_{ \{ X_{(i)} \}_{i \in \mathcal{I}}  } \, \Big \|  \, \prod_{i \in \mathcal{I} }  P_{ X_{(i)} }  \Big) \!\le\!  D_f   \Big(  P_{ \{ U_{(i)} \}_{i \in \mathcal{I}}  } \, \Big\|  \, \prod_{i \in \mathcal{I} }  P_{ U_{(i)} } \Big),
\end{equation} 
\item  (Mutual Information). For $\mathcal{I}_1, \mathcal{I}_2 \subset [n]$,
\begin{equation}
I(X_{(\mathcal{I}_1)}; X_{(\mathcal{I}_2)})  \le  I(U_{(\mathcal{I}_1)}; U_{(\mathcal{I}_2)}).  \label{eq:Bound_MI_Arbitrarry}
\end{equation} 
\end{itemize}

\end{theorem} 

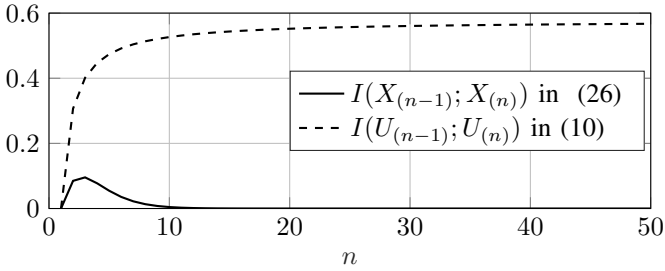
\begin{figure}[h]
\center
%
%
\definecolor{mycolor1}{rgb}{0.00000,0.44700,0.74100}%
\definecolor{mycolor2}{rgb}{0.85000,0.32500,0.09800}%
\begin{tikzpicture}

\begin{axis}[%
width=8cm,
height=2.6cm,
at={(1.011in,0.642in)},
scale only axis,
xmin=0,
xmax=50,
xlabel style={font=\color{white!15!black}},
xlabel={$n$},
ymin=0,
ymax=0.6,
axis background/.style={fill=white},
xmajorgrids,
ymajorgrids,
legend style={legend cell align=left, align=left, draw=white!15!black,at={(0.4,0.5)},anchor=west}
]
\addplot [color=black,thick]
  table[row sep=crcr]{%
1	0\\
2	0.0849495183976987\\
3	0.0956025889470326\\
4	0.0774159013507756\\
5	0.054580571792309\\
6	0.0356283647552121\\
7	0.0221402539954563\\
8	0.0132958316749298\\
9	0.00778616736761818\\
10	0.00447264286570748\\
11	0.00253044092241908\\
12	0.00141410699143739\\
13	0.000782274333743374\\
14	0.000429082007307156\\
15	0.000233658627629136\\
16	0.000126451720678812\\
17	6.80651356631227e-05\\
18	3.64646849849014e-05\\
19	1.94539844704494e-05\\
20	1.03403257496344e-05\\
21	5.47796619970748e-06\\
22	2.89340135112332e-06\\
23	1.52414561883138e-06\\
24	8.00903356688144e-07\\
25	4.19916663534759e-07\\
26	2.19713766776765e-07\\
27	1.14745221742644e-07\\
28	5.98217109791506e-08\\
29	3.11377008504933e-08\\
30	1.61833442060524e-08\\
31	8.39941997365472e-09\\
32	4.3538187487904e-09\\
33	2.25407404938572e-09\\
34	1.16567126380808e-09\\
35	6.02177218745461e-10\\
36	3.10770956518825e-10\\
37	1.60232116315024e-10\\
38	8.25419655732509e-11\\
39	4.24851642971296e-11\\
40	2.18502561002217e-11\\
41	1.12292423954988e-11\\
42	5.76681044926673e-12\\
43	2.95956281735211e-12\\
44	1.5178902570697e-12\\
45	7.78013910757054e-13\\
46	3.98548213295883e-13\\
47	2.04048022873674e-13\\
48	1.04412544683311e-13\\
49	5.34012943792751e-14\\
50	2.72985208042655e-14\\
};
\addlegendentry{$I(X_{(n-1)}; X_{(n)} )$ in~ \eqref{eq:MI_Bernoulli} }

\addplot [color=black, dashed, thick]
  table[row sep=crcr]{%
1	0\\
2	0.306852819440055\\
3	0.40138771133189\\
4	0.447038972213443\\
5	0.473895420899233\\
6	0.49157386410528\\
7	0.504089850944688\\
8	0.513415601177305\\
9	0.520632565520927\\
10	0.526383160974206\\
11	0.531072981169883\\
12	0.534970695089346\\
13	0.53826132074914\\
14	0.541076425518504\\
15	0.543512125460104\\
16	0.54564027098921\\
17	0.547515649172787\\
18	0.549180764744584\\
19	0.550669099029882\\
20	0.552007383589689\\
21	0.553217219420269\\
22	0.554316251404394\\
23	0.555319034288125\\
24	0.55623768073891\\
25	0.55708235288531\\
26	0.557861639732003\\
27	0.558582850210712\\
28	0.559252243076884\\
29	0.559875208979904\\
30	0.560456415924904\\
31	0.560999926435244\\
32	0.561509292636785\\
33	0.561987633970034\\
34	0.5624377011234\\
35	0.562861928956011\\
36	0.563262480560752\\
37	0.563641284150407\\
38	0.564000064095282\\
39	0.564340367165741\\
40	0.564663584822426\\
41	0.564970972232047\\
42	0.565263664555445\\
43	0.565542690954786\\
44	0.565808986683578\\
45	0.566063403558744\\
46	0.566306719062254\\
47	0.566539644276105\\
48	0.566762830822853\\
49	0.566976876953476\\
50	0.567182332901297\\
};
\addlegendentry{$I(U_{(n-1)}; U_{(n)})$ in~\eqref{eq:Exact_MI_Continious}}

\end{axis}

\end{tikzpicture}%
\vspace{-8mm}
\caption{$I(X_{(n-1)}; X_{(n)} )$ in~\eqref{eq:MI_Bernoulli} and $I(U_{(n-1)}; U_{(n)})$ in~\eqref{eq:Exact_MI_Continious} versus~$n$.}
\label{fig:MI_vs_n_Bernouilli_BOund}
\vspace{-0.5cm}
\end{figure}

\begin{IEEEproof}  
The proof relies on the data processing inequality.  Due to space constraints we only show it for $I( X_{(r)};  X_{(m)})$.

We start by defining the quantile function as
$
F_X^{-1}(y)= \sup \{ x : F_X(x) \le y  \}. 
$
Then, as discussed in the proof of Theorem~\ref{thm:f-diverngece}, for an arbitrary sampling distribution~\cite[eq.\ (1.1.3)]{arnold1992first},
\begin{equation}
(X_{(r)}, X_{(m)}) \stackrel{D}{=} ( F_X^{-1}(U_{(r)}),  F_X^{-1}(U_{(m)})),\label{eq:represenation_F_inverse}
\end{equation} 
thus  $I( F_X^{-1}(U_{(r)});  F_X^{-1}(U_{(m)})) = I( X_{(r)};  X_{(m)}).$ Moreover, 
\begin{align*}
I(U_{(r)}; U_{(m)})  &\ge I( F_X^{-1}(U_{(r)});  F_X^{-1}(U_{(m)})) 
= I( X_{(r)};  X_{(m)}), 
\end{align*} 
where the inequality uses the data processing inequality for the mutual information since $U_{(r)} \to U_{(m)} \to  F_X^{-1}(U_{(m)})$ and $ F^{-1}(U_{(r)}) \to U_{(r)} \to F_X^{-1}(U_{(m)})$ are Markov chains. This concludes the proof of Theorem~\ref{thm:UpperBOund}.
\end{IEEEproof}  
The bound in~\eqref{eq:Bound_MI_Arbitrarry} is appealing since all the results derived in Section~\ref{sec:Cont_Dist} (e.g., Theorem~\ref{thm:Limits_Continious_Case}) can be used to obtain upper bounds on $I(X_{(r)}; X_{(m)})$ for any arbitrary sampling distribution. 
 However, the bound in~\eqref{eq:Bound_MI_Arbitrarry} can be very suboptimal.
For example, as shown in Fig.~\ref{fig:MI_vs_n_Bernouilli_BOund},  we have that $\lim_{n \to \infty } I(X_{(n-1)}; X_{(n)} )  = 0$ for the Bernoulli case, while  $\lim_{n \to \infty }I(U_{(n-1)}; U_{(n)}) = \gamma$, computed in~\eqref{eq:1step}.

\begin{remark} Theorem~\ref{thm:UpperBOund} can be generalized to arbitrary non-overleaping subset of order statistics. Moreover, it can be generalized to $f$-divergences. 
\end{remark}

\begin{appendices}
\section{Proof of Proposition~\ref{prop:KL_div}}
\label{sec:Computation_Of_KL}

The computation for an arbitrary $\mathcal{I} \subseteq [n]$ with $|\mathcal{I}| = k$ proceeds as follows. First, notice
\begin{equation}
\begin{split}
&D_{\text{KL}} \Big(  P_{ \{ U_{(i)} \}_{i \in \mathcal{I}}  } \, \Big \|  \, \prod_{i \in \mathcal{I} }  P_{ U_{(i)} } \Big) \notag\\
& \stackrel{{\rm{(a)}}}{=} \E \left[  \log \left( \frac{c_{\mathcal{I}} 
 \prod_{t=1}^{k+1} \left [U_{(i_t)} - U_{(i_{t-1})} \right ]^{i_t-i_{t-1}-1}}{     \prod_{t=1}^{k}      c_{i_t}   \, U_{(i_t)}^{i_t-1} \, (1-U_{(i_t)})^{n-i_t}}   \right) \right],
 \label{eq:CombinginSums0}
 \end{split}
 \end{equation}
using the expression of the joint distribution in~\eqref{eq:Joint_Distribution} and 
$
P_{U_{(r)}}(x)=c_r   \, x^{r-1} \, (1-x)^{n-r},
$
where $c_r= \frac{n!}{(r-1)! (n-r)!}$. Then, we simplify further:
\begin{align}
& D_{\text{KL}} \Big(  P_{ \{ U_{(i)} \}_{i \in \mathcal{I}}  } \, \Big \|  \, \prod_{i \in \mathcal{I} }  P_{ U_{(i)} } \Big) \notag \\
 &=  \log \Big( c_{\mathcal{I}}   \prod_{t=1}^{k}      c_{i_t}^{-1}  \Big)   -   \sum_{t=1}^{k} (i_t-1) \E[ \log (  U_{(i_t)}  )    ] \nonumber \\
 & \quad -   \sum_{t=1}^{k} (n-i_t) \E[ \log (1-  U_{(i_t)}  )    ]\nonumber \\
  &  \quad + \sum_{t=1}^{k+1}  (i_t-i_{t-1}-1) \E[ \log (  U_{(i_t)} - U_{(i_{t-1})}   )    ]\nonumber \\
 &=\log \Big( c_{\mathcal{I}}   \prod_{t=1}^{k}      c_{i_t}^{-1}  \Big) -   \sum_{t=1}^{k} (i_t-1) (\psi(i_t) - \psi(n+1)) \nonumber \\
  & \quad -   \sum_{t=1}^{k} (n-i_t) (\psi( n +1-i_t ) - \psi(n+1))  \nonumber \\
 &  \quad + \sum_{t=1}^{k+1}  (i_t-i_{t-1}-1) (  \psi(i_t-i_{t-1}   ) - \psi(n+1 ) ),
 \label{eq:CombinginSums1}
 \end{align}
 where we have used the fact that  $U_{(m)} \sim {\rm Beta}( m,   n +1-m )$ and $1-U_{(r)}  \sim {\rm Beta}(   n +1-r,r ) $ with the difference $U_{(m)}-U_{(r)}  \sim {\rm Beta}( m-r,   n -(m-r)+1 )$. Then, (see, e.g.,~\cite{arnold1992first}),
\begin{align*}
\E \left[ \log  (U_{(m)} )  \right] &= \psi(m ) - \psi(n+1),  \\
\E \left[ \log  (1-U_{(r)} )  \right] &=     \psi( n +1-r ) - \psi(n+1),\\
 \E \left[ \log  (U_{(m)}-U_{(r)} )  \right]&=  \psi(m-r   ) - \psi(n+1 ) ,
\end{align*}
where $\psi(\cdot)$ is the digamma function and where we use the convention that $U_{(0)}=0$ and $U_{(n+1)}=1$. 
 Finally, collecting terms and using the result of \eqref{eq:CombinginSums1}, we have
\begin{align}
 &D_{\text{KL}} \Big(  P_{ \{ U_{(i)} \}_{i \in \mathcal{I}}  } \, \Big \|  \, \prod_{i \in \mathcal{I} }  P_{ U_{(i)} } \Big)\nonumber \\
  &=\log \Big( c_{\mathcal{I}}   \prod_{t=1}^{k}      c_{i_t}^{-1}  \Big)+   \sum_{t=1}^{k+1}  (i_t-i_{t-1}-1)  \psi(i_t-i_{t-1}   ) \nonumber\\
 & \quad -   \sum_{t=1}^{k} (i_t-1) \psi(i_t)  -   \sum_{t=1}^{k} (n-i_t) \psi( n +1-i_t )\nonumber   \\
 & \quad + \psi(n+1 ) \left[\sum_{t=1}^{k} [(i_t-1) + (n-i_t)]  - \sum_{t=1}^{k+1}  (i_t-i_{t-1}-1)  \right]\nonumber \\
 &\stackrel{{\rm{(a)}}}{=}  \log \Big( c_{\mathcal{I}}   \prod_{t=1}^{k}      c_{i_t}^{-1}  \Big) +  \sum_{t=1}^{k+1}  (i_t-i_{t-1}-1)  \psi(i_t-i_{t-1}   ) \nonumber\\
 & \quad -   \sum_{t=1}^{k} (i_t-1) \psi(i_t)  -   \sum_{t=1}^{k} (n-i_t) \psi( n +1-i_t )\nonumber \\
 & \quad + \psi(n+1 ) \left(k(n-1) -(n-k)   \right) \nonumber \\
& \stackrel{{\rm{(b)}}}{=} \!-\! \sum_{t=1}^{k+1}\!T_{i_t-i_{t-1}-1} \!+\! \sum_{t=1}^k \!T_{i_t-1}\!+\!\sum_{t=1}^k \!T_{n-i_t}\!-\!(k\!-\!1)T_n,
\label{eq:CombinginSums2}
 \end{align}
where the labeled equalities follow from:
$\rm{(a)}$
 the fact that
\begin{align*}
\sum_{t=1}^{k+1}  (i_t-i_{t-1}-1)= i_{k+1}-i_0-(k+1) = n-k,
\end{align*}
where we recall that $i_0=0$ and $i_{k+1}=n+1$;
and $\rm{(b)}$ using the identity $\psi(n)= H_{n-1}-\gamma$, with $\gamma $ being the Euler-Mascheroni constant, noting that the terms that multiply $\gamma$ cancel out, and using the definition of $T_n$ from~\eqref{eq:Tr_def}. The result in \eqref{eq:KL_expression} comes from canceling the terms $T_{i_1 - i_0-1} = T_{i_1-1}$ and $T_{i_{k+1}-i_{k}-1} = T_{n-i_{k}}$ from \eqref{eq:CombinginSums2}.

The special case of   $\mathcal{I}=[n]$  follows by observing that $i_t=t$ for $t \in \{0,1,...,n+1\}$ and hence $i_t-i_{t-1}=1$ leading to
\begin{align*}
&D_{\text{KL}} \Big(  P_{ \left \{U_{(1)}, ...,  U_{(n)} \right \} } \, \Big\|  \, \prod_{i=1 }^n  P_{ U_{(i)} }\Big)\\
&=  \sum_{t=2}^n (T_{i_t-1} -T_{i_t-i_{t-1}-1})+\sum_{t=1}^{n-1} T_{n-i_t}-(n-1)T_n  \\
&=  \sum_{t=2}^n (T_{t-1} -T_{t-(t-1)-1})+\sum_{t=1}^{n-1} T_{n-t}-(n-1)T_n  \\
&=  2\sum_{t=2}^n T_{t-1} -(n-1)T_n,
\end{align*}
where the first equality follows from~\eqref{eq:KL_expression}, and 
the last equality follows since $T_0=0$ and noting that $ \sum_{t=2}^n T_{t-1}=\sum_{t=1}^{n-1} T_{n-t}$.

The special case of $\mathcal{I}=\{ 1, n\}$ (hence $k=2$ with $i_2 = n$ and $i_1=1$)  follows by observing that \eqref{eq:KL_expression} gives
\begin{align*}
&D_{\text{KL}} \big(  P_{ \left \{U_{(1)}, \,  U_{(n)} \right \}} \,  \big\|  \,   P_{ U_{(1)} }  P_{ U_{(n)} }  \big)  
 = T_{n-1} -T_{n-2} + T_{n-1}  -T_n
\\& = H_{n-1} - H_{n-2} - \log(n) + \log(n-1)
\\& = \log \left( \frac{n-1}{n} \right ) + \frac{1}{n-1},
\end{align*}
and the second equality follows since
\begin{align}
T_{k+1} - T_{k} &=  \!\log((k+1)!)- (k+1)H_{k+1} -  \log(k!) + kH_{k} \nonumber \\
& =  \log(k+1) - H_{k} -1,
\label{eq:step1Diff}
\end{align}
where the last equality follows since $(k+1)H_{k+1} = (k+1)H_{k} + 1$ by the definition of the harmonic mean in~\eqref{eq:Hr_def}.

\section{Proof of \eqref{eq:cov_1_2_exponential}}\label{app:eq:cov_1_2_exponential}
First, by~\cite[page~18]{David2003Book}, if the sample is drawn from an exponential distribution then the $k^{th}$ order statistic for $k \in [n]$ is distributed as
$
X_{(k)} \stackrel{D}{=}\frac{1}{\lambda} \sum_{i=1}^k \, a_{n,i} \, Z_i,
$
where the $Z_i$'s are i.i.d.\ exponential random variables with rate one and $a_{n,i} = [n-i+1]^{-1}$.  
Then,
$\E[X_{(k)}] = \frac{1}{\lambda} \sum_{i=1}^k \, a_{n,i} \, \E[Z_i] =  \frac{1}{\lambda} \sum_{i=1}^k a_{n,i},$ since $\E[Z_i] = 1$.
Thus, $\E[X_{(1)}] \!=\!  { a_{n,1}}/{\lambda}$ 
and $\E[X_{(2)}] \!=\!  { a_{n,1}}/{\lambda} \!+\!  {a_{n,2}}/{\lambda}$. 
Moreover, using $\E[Z_i^2] = 2$,
\begin{align*}
&\E[X_{(1)} X_{(2)}] =\frac{1}{\lambda^2} \E [( a_{n,1} Z_1) ( a_{n,1} Z_1 +  a_{n,2} Z_2)] \\
&=\frac{1}{\lambda^2}( a_{n,1}^2 \E [ Z^2_1] + a_{n,1} a_{n,2}  \E[Z_1 Z_2] ) =\frac{1}{\lambda^2}( 2a_{n,1}^2 + a_{n,1} a_{n,2}).
\end{align*}
Therefore, using the fact that $a_{n,1} = {1}/{n}$,
\begin{align*}
&{\rm Cov}(X_{(1)}, X_{(2)}) = \E[ X_{(1)} X_{(2)}]-  \E[ X_{(1)}] \E[ X_{(2)}]\\
&= \frac{1}{\lambda^2}\left( 2a_{n,1}^2 + a_{n,1} a_{n,2}  \right) - \frac{1}{\lambda^2}\left( a^2_{n,1} +  a_{n,1} a_{n,2}\right) \\
&= \frac{1}{\lambda^2} a_{n,1}^2 = \frac{1}{\lambda^2 n^2}.
\end{align*}


\section{Proof of Lemma~\ref{lem:approx_lemma}} \label{app:lem:approx_lemma}
By~\cite{detemple199175} (see also~\cite[page 76]{Havil2003}), 
\begin{equation}
kH_k -k \log\left(k+  \frac{1}{2} \right)- k \gamma= e_H(k), \label{eq:Approximation_Harmonic_Number}
\end{equation} 
where $  \frac{k}{24 (k+1)^2}  \le   e_H(k) \le    \frac{1}{24 k}$,
and using a strong form of Stirling's approximation \cite{robbins1955remark}, for $\frac{1}{12k+1}  \le   e_f(k) \le    \frac{1}{12 k}$,
\begin{equation}
\log(k!)- \left( \frac{1}{2}\log(2\pi k) +k \log(k)-k\right)= e_f(k). \label{eq:Approximation_log_factorial} 
\end{equation} 
Combining~\eqref{eq:Approximation_Harmonic_Number} and~\eqref{eq:Approximation_log_factorial} with the definition of $T_k$ in~\eqref{eq:Tr_def}, namely $T_k = \log(k!)- kH_k$, we find the result in \eqref{eq:Tk_approx},
where we have defined $e(k) := e_H(k)-e_f(k)$ and using the bounds for $e_H(k)$ and $e_f(k)$ stated above, we find the result in \eqref{eq:e_approx}.

Next, we show~\eqref{eq:diff_of_T(n)-T(n-1)}. First, from the definition of $T_k$ in~\eqref{eq:Tr_def},
we have that~\eqref{eq:step1Diff} holds.
Next, observe that by  \eqref{eq:Approximation_Harmonic_Number},
\begin{equation}
\label{eq:Hkmin1}
H_k = H_{k+1} - \frac{1}{k+1} = \log\left(k+  \frac{3}{2} \right)+\gamma +\frac{e_H(k+1)-1}{k+1} ,
\end{equation} 
and therefore, by substituting this inside~\eqref{eq:step1Diff}, we obtain \eqref{eq:diff_of_T(n)-T(n-1)} with
$c(k)$ defined as $c(k):={e_H(k+1)}/{(k+1)}$ and the bounds  in \eqref{eq:ck} follow from $  \frac{k+1}{24 (k+2)^2}  \le   e_H(k+1) \le    \frac{1}{24 (k+1)}$.


\section{Proof of Theorem~\ref{thm:Limits_Continious_Case}} \label{app:thm:Limits_Continious_Case}
We consider each case separately.
 \subsection{Case 1: Proof of the $r^{th}$ vs. Max} 
We set $m=n$ in~\eqref{eq:Exact_MI_Continious} and  use~\eqref{eq:diff_of_T(n)-T(n-1)}. With this, we obtain
 \begin{align*}
& I(X_{(r)}; X_{(n)}) = -(T_{n} -T_{n-1}) + (T_{n-r} - T_{n-r-1}) \notag\\
&= \log \left(\frac{2n+1}{2n} \right) -\log \left(\frac{2(n-r)+1}{2(n-r)}\right) -\frac{1}{n} +\frac{1}{n-r} \\
&\qquad +c(n-1) -c(n-r-1). 
 \end{align*} 
Finally, taking the limit of the above we find result~\eqref{eq:r-th_order_vs_max} 
using the limits in Table~\ref{table:PropLimCase1}.
Specifically, the Table~\ref{table:PropLimCase1} limits follow from:
(\#1) the Maclaurin series for the natural logarithm;
(\#2) combining the two terms inside the bracket and taking the limit;
and (\#3) and (\#4) from the bounds in \eqref{eq:ck}, namely $\frac{1}{24 (k+2)^2}  \le  c(k) \le   \frac{1}{24 (k+1)^2}$.
\begin{table}[ht]
\centering
\begin{tabular}[t]{lcc}
\hline
& function $f(n)$ & $\lim_{n \to \infty } f(n)$\\
\hline
(\#1) & $n^2 \left( \log \left(\frac{2(n-r)}{2(n-r)+1}\right) - \log \left(\frac{2n}{2n+1}\right) \right)$ & $-\frac{r}{2}$\\
(\#2) & $n^2 \left( \frac{1}{n-r} - \frac{1}{n} \right)$ & $r$\\
(\#3) & $n^2c(n-r-1)$ & $1/24$\\
(\#4) & $n^2 c(n-1)$ & $1/24$\\
\hline
\end{tabular}
\vspace{2mm}
\caption{List of limits for Case~1.}
\label{table:PropLimCase1}
\vspace{-5mm}
\end{table}

\subsection{Case 2: Proof of the $r^{th}$ vs. the $m^{th}$} 
Using~\eqref{eq:Exact_MI_Continious}, we have that 
\begin{align*}
\lim_{n \to \infty} I(X_{(r)}; X_{(m)}) &= \lim_{n \to \infty} [T_{m-1} + T_{n-r} - T_{m-r-1}-T_{n}] \notag \\
&=  T_{m-1} - T_{m-r-1}+ (1+\gamma) r, 
\end{align*} 
where we have used the fact that, from~\eqref{eq:Tk_approx}, we have
\begin{align}
&T_{n-r} -T_{n}
=  (n-r) \log \hspace{-0.05cm} \left( \hspace{-0.05cm}  \frac{2(n-r)}{2(n-r)+1} \hspace{-0.05cm} \right) -  n \log \hspace{-0.05cm}  \left(\hspace{-0.05cm}  \frac{2n}{2n+1} \hspace{-0.05cm}  \right)   \notag\\
&\qquad  + \frac{1}{2} \log \left( \frac{n-r}{n} \right)  +(1+\gamma) r - e(n-r) + e(n) , \label{eq:simplified_diff}
\end{align} 
and hence,
\begin{equation}
   \lim_{n \to \infty} \left(T_{n-r} -T_{n} \right)=(1+\gamma) r,  
\label{eq:Difference_of_Ts}
\end{equation}
which can be seen by expanding with the Maclaurin series for the natural logarithm,  for example, we have,
\begin{align*}
(n-r) \log \left(\frac{2(n-r)}{2(n-r)+1}\right) = \sum_{i=1}^{\infty} \frac{n-r}{i} \left(\frac{-1}{2(n-r)+1}\right)^i ,
\end{align*}
and hence the limit equals $=  -{1}/{2}$, along with~\eqref{eq:e_approx} from which it follows that $ \lim_{n \to \infty} e(n-r)  =  \lim_{n \to \infty} e(n) = 0$.

\subsection{Case 3: Proof of the $k$-Step}
To show \eqref{eq:k-step_limit} observe that 
\begin{align*}
 & \lim_{n \to \infty}  I(X_{(n-k)}; X_{(n)}) \stackrel{{\rm{(a)}}}{=}    \lim_{n \to \infty} [T_{n-1} + T_{k} - T_{k-1}-T_{n]} \\
   &\stackrel{{\rm{(b)}}}{=}   T_{k} - T_{k-1} + (1+\gamma) 
   \stackrel{{\rm{(c)}}}{=} \log(k)  -H_{k-1} +\gamma\\
   &\stackrel{{\rm{(d)}}}{=}   \log \left(\frac{2k}{2k+1}\right)+\frac{1}{k} -c(k-1), 
\end{align*}
where the labeled equalities follow from: $\rm{(a)}$ eq.\ \eqref{eq:Exact_MI_Continious}; $\rm{(b)}$ eq.\  \eqref{eq:Difference_of_Ts};
$\rm{(c)}$ eq.\  \eqref{eq:step1Diff}; and $\rm{(d)}$ eq.\  \eqref{eq:Hkmin1} and $c(k-1) = e_H{(k)}/k$.

\subsection{Case 4: Proof of $\lfloor \alpha n \rfloor$ vs. $\lceil \beta n \rceil $ }
\begin{table}[ht]
\centering
\caption{List of limits for Case~4.}
\label{table:PropLimCase4}
\vspace{-0.5cm}
\begin{tabular}[t]{lcc}
\hline
&  function $f(n)$ & $\lim_{n \to \infty } f(n)$\\
\hline
(\#1) & $(\lceil \beta n \rceil -\lfloor\alpha n \rfloor - 1) \log \left( \frac{2(\lceil \beta n \rceil -\lfloor\alpha n \rfloor -1)}{2(\lceil \beta n \rceil -\lfloor\alpha n \rfloor)-1}\right)$ & $-\frac{1}{2}$\\
(\#2) & $(\lceil \beta n \rceil -1) \log \left(\frac{2(\lceil \beta n \rceil -1)}{2\lceil \beta n \rceil -1} \right )$ & $-\frac{1}{2}$\\
(\#3) & $(n-\lfloor\alpha n \rfloor) \log \left( \frac{2(n-\lfloor\alpha n \rfloor)}{2(n-\lfloor\alpha n \rfloor)+1} \right )$ & $-\frac{1}{2}$\\
(\#4) & $n \log \left( \frac{2n}{2n+1} \right )$ & $-\frac{1}{2}$ \\
(\#5) & $\frac{1}{2} \log \left( \frac{\lceil \beta n \rceil -\lfloor\alpha n \rfloor-1}{\lceil \beta n \rceil -1} \right)$ & $\frac{1}{2} \log \left( \frac{\beta-\alpha}{\beta}\right)$\\
(\#6) & $ \frac{1}{2} \log \left( \frac{n-\lfloor\alpha n \rfloor}{n} \right)$ & $\frac{1}{2} \log \left(1-\alpha \right)$\\
\hline
\end{tabular}
\end{table}
Setting $r=\lfloor \alpha n \rfloor$ and $m=\lceil \beta n \rceil$ in~\eqref{eq:Exact_MI_Continious} and then using~\eqref{eq:simplified_diff} and~\eqref{eq:Tk_approx}, we obtain
\begin{align*}
&I(X_{(\lfloor \alpha n \rfloor)}; X_{(\lceil \beta n \rceil )}) \notag\\
&=T_{\lceil \beta n \rceil -1} + T_{n-\lfloor\alpha n \rfloor} - T_{\lceil \beta n \rceil -\lfloor\alpha n \rfloor -1}
-T_{n} \notag\\
&=  -(\lceil \beta n \rceil-\lfloor\alpha n \rfloor -1) \log \left( \frac{2(\lceil \beta n \rceil -\lfloor\alpha n \rfloor -1)}{2(\lceil \beta n \rceil -\lfloor\alpha n \rfloor)-1} \right )  \notag\\
&\quad - \frac{1}{2} \log \left( \frac{\lceil \beta n \rceil -\lfloor\alpha n \rfloor  -1}{\lceil \beta n \rceil -1} \right)  + e(\lceil \beta n \rceil - \lfloor\alpha n \rfloor  -1) \notag\\
& \quad  + (\lceil \beta n \rceil -1) \log \left( \frac{2(\lceil \beta n \rceil -1)}{2\lceil \beta n \rceil -1}   \right ) - e(\lceil \beta n \rceil-1 ) \notag\\
& \quad + (n-\lfloor\alpha n \rfloor) \log \left( \frac{2(n-\lfloor\alpha n \rfloor)}{2(n-\lfloor\alpha n \rfloor) +1} \right ) + \frac{1}{2} \log \left( \frac{n-\lfloor\alpha n \rfloor}{n} \right)  \notag\\
& \quad - e(n-\lfloor\alpha n \rfloor) -  n \log \left( \frac{2n}{2n+1} \right ) + e(n), 
\end{align*} 
and then using the limits in Table~\ref{table:PropLimCase4}, and the fact that $ \lim_{k \to \infty} e(k)  =  0$  from~\eqref{eq:e_approx},
\begin{equation}
\lim_{n \to \infty}   I(X_{(\lfloor \alpha n \rfloor)}; X_{(\lceil \beta n \rceil )})= \frac{1}{2} \log \left( \frac{\beta (1-\alpha)}{\beta-\alpha} \right).
\end{equation} 
Specifically, the limits in Table~\ref{table:PropLimCase4} follow from:
(\#1)-(\#4) the Maclaurin series for the natural logarithm, and (\#5)-(\#6) the fact that, for fixed $0 < \alpha < \beta <1$ with $(\alpha,\beta)$ independent of $n$, we can write $ \lfloor \alpha n \rfloor = \alpha n - \{\alpha n\}$ and $ \lceil \beta n \rceil = \beta n + \{\beta n\}$, where $\{x\}$ indicates the fractional part of $x$, i.e, $0 \leq \{x\} <1$ with $x \in \{\alpha n,\beta n\}$.

\subsection{Case~5: Proof of $\lfloor \alpha n \rfloor$ vs. Max}
We set $r = \lfloor \alpha n \rfloor$ and $m=n$ in~\eqref{eq:Exact_MI_Continious} and we use~\eqref{eq:diff_of_T(n)-T(n-1)}. With this, we obtain
\begin{align*}
&  I(X_{(\lfloor \alpha n \rfloor)}; X_{(n)}) = T_{n-1} + T_{n-\lfloor \alpha n \rfloor} - T_{n-\lfloor \alpha n \rfloor-1}-T_{n}\\
&= \! -\log \left( \frac{2n}{2n+1} \right )\!-\!\frac{1}{n} + c(n-1) - c(n-\lfloor \alpha n \rfloor-1) \notag\\
&  \quad +\log \left( \frac{2(n-\lfloor \alpha n \rfloor)}{2(n-\lfloor \alpha n \rfloor)+1} \right )+\frac{1}{n-\lfloor \alpha n \rfloor}.
\end{align*}  
Finally, the limit in~\eqref{eq:alpha_max_limit_n} is given by
\begin{align*}
\lim_{n \to \infty}  n I(X_{(\lfloor \alpha n \rfloor)}; X_{(n)})
&=  \frac{1}{2}-1 - \frac{1}{2 (1-\alpha)}+\frac{1}{1-\alpha} \\
&= \frac{\alpha}{2(1-\alpha)},
\end{align*}
where we have used the limit (\#4) in Table~\ref{table:PropLimCase4}, 
the fact that from~\eqref{eq:ck} we have $\lim_{n \to \infty} n c(n) = \lim_{n\to \infty} n c(n-\lfloor \alpha n \rfloor) = 0$,
and the facts that 
\begin{align*}
\lim_{n \to \infty} n \log \left( \frac{2(n-\lfloor \alpha n \rfloor)}{2(n-\lfloor \alpha n \rfloor)+1} \right ) &= - \frac{1}{2(1-\alpha)},
\\
\lim_{n \to \infty} \frac{n}{n-\lfloor \alpha n \rfloor} &= \frac{1}{1-\alpha},
\end{align*}
which follow by using the Maclaurin series for the natural logarithm and since, for fixed $0 < \alpha <1$ with $\alpha$ independent of $n$, we can write $ \lfloor \alpha n \rfloor = \alpha n - \{\alpha n\}$, where $0 \leq \{\alpha n\} <1$.

\section{Proof of Theorem~\ref{thm:Bernoulli_Case}} 
\label{app:thm:Bernoulli_Case}
First, notice that $P(X_{(k)}=0) $ is the probability that there are $k$ or more zeros in the sample, hence, for $B$ a ${\rm Binomial}(n, 1-p)$ random variable, $P(X_{(k)}=0) = P( B \ge k  )$ and $P(X_{(k)}=1)=1- P( B \ge k  ).$ Moreover, we observe that $P(X_{(r)}=0, X_{(m)}=0) =  P(X_{(m)}=0)=P( B \ge m  )$ and $P(X_{(r)}=1, X_{(m)}=1) 
=P(X_{(r)}=1)=1- P( B \ge r  )$. Finally, we will use that
\begin{align*}
&P(X_{(r)}=0, X_{(m)}=1)   \\
&=  P(X_{(r)}=0) - P(X_{(r)}=0, X_{(m)}=0)  \\
&  =  P(X_{(r)}=0) - P(X_{(m)}=0)  = P( B \ge r) - P( B \ge m).
\end{align*}
Therefore, the mutual information is given by 
\begin{align*}
&I(X_{(r)}; X_{(m)} ) \notag\\
&= P(X_{(r)}=0, X_{(m)}=0) \log  \left(\frac{P(X_{(r)}=0, X_{(m)}=0)}{ P(X_{(r)}=0)  P(X_{(m)}=0)} \right)\notag\\
&+ P(X_{(r)}=0, X_{(m)}=1) \log\left( \frac{P(X_{(r)}=0, X_{(m)}=1)}{ P(X_{(r)}=0)  P(X_{(m)}=1)} \right)\notag\\
&+ P(X_{(r)}=1, X_{(m)}=1) \log\left( \frac{P(X_{(r)}=1, X_{(m)}=1)}{ P(X_{(r)}=1)  P(X_{(m)}=1)} \right).
\end{align*} 
Plugging in the values for these probabilities in terms of $B$ as discussed above gives the result in \eqref{eq:MI_Bernoulli}.


\end{appendices} 

\bibliography{refs}
\bibliographystyle{IEEEtran}
\end{document}